\documentclass[journal=apchd5,manuscript=article]{achemso}

\usepackage[version=3]{mhchem} 
\usepackage[T1]{fontenc}       
\usepackage{SIunits}

\usepackage{ulem}
\usepackage{color}

\author{Andreas W. Schell}
\altaffiliation{These authors contributed equally}
\affiliation[HU Berlin]
{AG Nanooptik, Humboldt-Universit\"at zu Berlin, 12846 Berlin, Germany}
\alsoaffiliation[Kyoto University]
{Department of Electronic Science and Engineering, Kyoto University, 615-8510 Kyoto, Japan}
\email{schell@physik.hu-berlin.de}
\author{Alexander Kuhlicke}
\altaffiliation{These authors contributed equally}
\author{G\"unter Kewes}
\altaffiliation{These authors contributed equally}
\author{Oliver Benson}
\affiliation[HU Berlin]
{AG Nanooptik, Humboldt-Universit\"at zu Berlin, 12846 Berlin, Germany}

\title{'Flying Plasmons': Fabry-P\'{e}rot Resonances in Levitated Silver Nanowires}

\keywords{Paul trap, silver nanowires, surface plasmon polaritons, Fabry-P\'{e}rot resonances, deposition, AFM}

\begin{document}

\begin{abstract}
Plasmonic nano structures such as wire waveguides or antennas are key building blocks for novel highly integrated photonics. A quantitative understanding of the optical material properties of individual structures on the nanoscale is thus indispensable for predicting and designing the functionality of complex composite elements.  In this letter we study propagating surface plasmon polaritons in single silver nanowires isolated from its environment by levitation in a linear Paul trap. Symmetry-breaking effects, e.g., from supporting substrates are completely eliminated in this way. Illuminated with white light from a supercontinuum source, Fabry-P\'{e}rot-like resonances are observed in the scattering spectra obtained from the ends of the nanowires. The plasmonic nature of the signal is verified by local excitation and photon collection corresponding to a clean transmission measurement through a Fabry-P\'{e}rot resonator. A numerical simulation is used to compute the complex effective refractive indices of the nanowires as input parameter for a simple Fabry-P\'{e}rot model, which nicely reproduces the measured spectra despite the highly dispersive nature of the system. Our studies pave the way for quantitative characterization of nearly any trappable plasmonic nano object, even of fragile ones such as droplets of liquids or molten metal and of nearly any nanoresonator based on a finite waveguide with implications for modeling of complex hybrid structures featuring strong coupling or lasing. Moreover, the configuration has the potential to be complemented by gas sensors to study the impact of hot electrons on catalytic reactions nearby plasmonic particles.

\end{abstract}

The field of plasmonics is motivated by the ever increasing need to reduce the footprint of optical devices, to enhance their operation speed, and to increase their energy efficiency~\cite{Guo2013}. Surface plasmon polaritons (SPPs) at metal-dielectric interfaces can tightly confine and guide electromagnetic excitations~\cite{Barnes2003}. This not only allows for guiding energy on small length scales, but also establishes a dramatically enhanced light-matter interaction. In order to understand and even design the functionality of highly integrated devices utilizing SPPs, it is mandatory to exactly know the properties of individual constituents, i.e. plasmonic nano wires or nano particles. However, the material parameters of nano particles deviate from bulk properties and mandatory supporting structures are of great influence.
Along their application in photonics, (localized) SPPs are very attractive as local generators of heat~\cite{Brongersma2015,Baffou2013}, point-like electron emitters~\cite{Bormann2010, Krueger2011}, and local chemical reaction sites~\cite{Linic2015}. Here again, it would be desirable to work out the properties of a 'pure' plasmonic nano particle, e.g., in a comparison with theoretical investigations looking for signatures of quantum behavior~\cite{Tame2013}. 

A way to isolate small objects in a clean environment is levitation in vacuum. Optical levitation using focussed lasers has been studied extensively~\cite{Ashkin1976,Li2000,Gieseler2012,Gieseler2014a,Gieseler2014b}, but may induce heat due to absorption, in particular in metal objects. Here, we follow another approach: levitation by charged particles using the oscillating electric field in a Paul trap~\cite{Paul1990}. Particles ranging in size from a few micrometers~\cite{Kuhlicke2014} down to single ions~\cite{Cirac1995} can be trapped and investigated for arbitrarily long times. In addition to mere levitation, Paul traps can also be utilized for controlled assembly of composite structures~\cite{Kuhlicke2015} or potentially coupling to quantum emitters such as the nitrogen vacancy center in diamond~\cite{Kuhlicke2014,Delord2016} or graphene~\cite{Nagornykh2015}.

In this letter, we employ a Paul trap optimized for studying the optical properties of particles of about a micron in diameter~\cite{Kuhlicke2015}. We focus our study on an accepted model system for plasmonics, i.e. silver nanowires.  The optical properties of nanowires and their interaction with light emitters have been extensively studied~\cite{Ditlbacher2005,Akimov2007,Wiley2008,Kolesov2009,Schell2011,Huck2011,Rewitz2011,Wild2012,Yan2012,Ropp2013,Schell2014,Yi2017} with wires spin-coated or deposited on substrates observing Fabry-P\'{e}rot-like spetral responses. We want to point out though, that different interpretations of these spectral features have been discussed~\cite{Wild2012} and that most of the papers lack a thorough numerical analysis. Therefore we feel, that a revisit of this topic is overdue. In recent years also trapping of individual nanowires has been performed employing optical traps. For the first time, we are able to investigate silver nano wires freely levitated in a Paul trap. This allows for studies of their optical properties without any supporting structure and in an environment, which can be almost completely controlled. Furthermore, the simple and well-defined geometry of the problem, i.e. a cylinder surrounded by vacuum or air, renders precise, quantitative comparison to analytical and numerical models possible. This work represents a study of unprecedented coherence, as we are able to correlate clean transmission spectrograms with wire geometry and a simple Fabry-P\'{e}rot model.

\begin{figure}
	\centering
		\includegraphics[width=\linewidth]{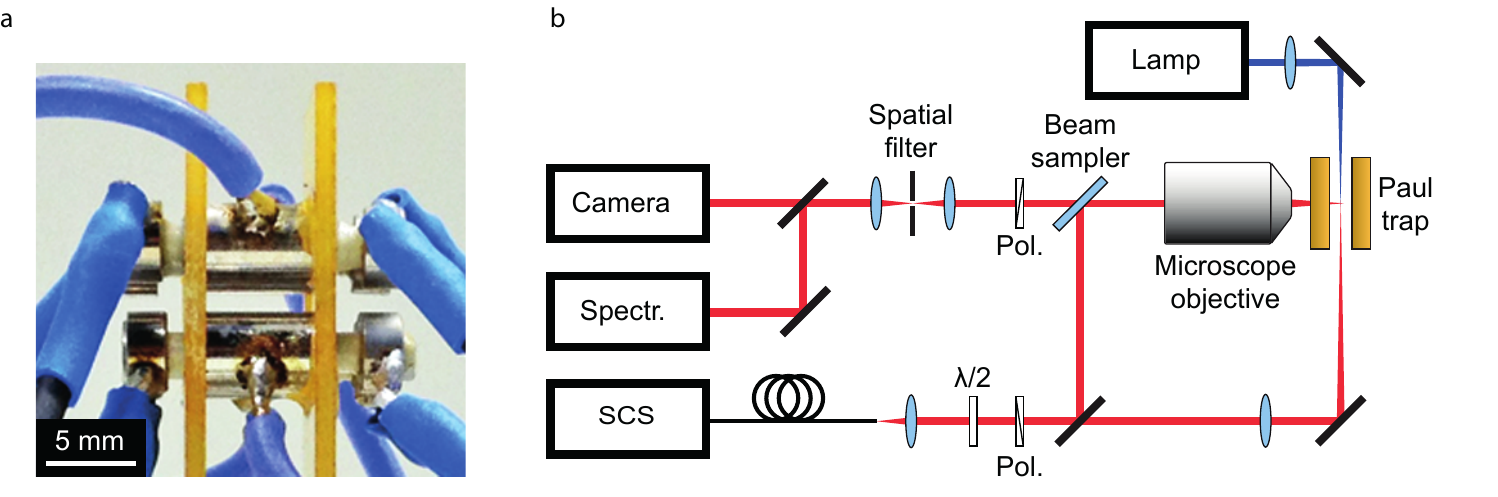}
		\caption{\textbf{Paul trap and microscope setup.}
		\textbf{a}, Photo of the linear Paul trap.
		\textbf{b}, Sketch of the setup used in the experiment. The wire in the trap can be illuminated using a halogen lamp, or alternatively by a supercontinuum source (SCS) shining either along the axis of the linear trap or focused through a microscope objective. Detection is performed through the objective lens by a CCD camera or a spectrometer. See methods section for more details.
		}
		\label{fig:setup}
\end{figure}

We use a special configuration of a Paul trap, a linear Paul trap (see Figure~\ref{fig:setup}(a)), to levitate charged silver nanowires in free space. The trap is integrated in a homemade optical microscope setup, as illustrated in Figure~\ref{fig:setup}(b), which is used to detect and characterize the trapped particles (more details can be found in the methods section). 
The geometry of the electrodes restricts the optical access, but with different trap designs, such as a needle trap, it is still possible to use high NA objective lenses.

Silver nanowires (PlasmaChem) with an average diameter of $\unit{100}{\nano\meter} \pm \unit{20}{\nano\meter}$ and lengths of up to \unit{50}{\micro\meter} are dispersed in ethanol and injected into the Paul trap using electrospray ionization~\cite{Fenn1989}. After injection, the axial trapping potential is adjusted by the dc voltages applied to the end-cap electrodes in order to isolate a single nanowire in front of the microscope objective. The trajectory of the confined particle is described by a micromotion driven with the trap frequency and a slower macromotion. The range of these oscillations can be reduced by adjusting the frequency and the amplitude of the ac trap voltage~\cite{Kuhlicke2014}. A limit to stabilization is imposed by geometric imperfections of the trapping field and coupling to micromotion through the nanowires' elongated shape as compared to an object with a point-like mass and charge distribution.

\begin{figure}
	\centering
		\includegraphics[width=0.7\linewidth]{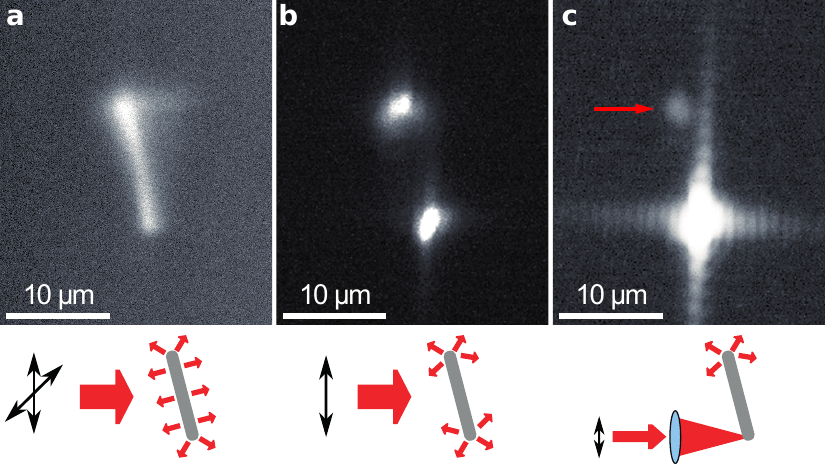}
		\caption{\textbf{Optical characterization of a single levitated silver nanowire.}
		\textbf{a}, CCD camera image of a nanowire illuminated by unpolarized white light from a halogen lamp. With this illumination, the whole wire can be seen.
		\textbf{b}, Using  white light from the supercontinuum source polarized parallel to the wire  strong scattering from 
		the wire's ends is visible.
		\textbf{c}, White light from the supercontinuum source is focussed onto the lower end of the wire. Besides the large bright spot from the directly reflected light, light also stemming from the distal end of the wire is visible (indicated by an arrow). This is a clear indication of SPPs being guided along the wire and scattered out at the end. The length of the wire is approximately \unit{12}{\micro\meter}.	
		}
		\label{fig:opt}
\end{figure}

A first optical characterization is carried out for individual silver nanowires trapped in front of the microscope objective. Figure~\ref{fig:opt}(a) shows a silver nanowire of \unit{12}{\micro\meter} in length illuminated by white light from the halogen lamp. Most of the trapped nanowires align in vertical direction due to an asymmetry of the trapping potential caused by gravity and the difference of trapping strength in the radial and the axial trap direction. Figure~\ref{fig:opt}(b) shows the same wire illuminated from the side by white light from a supercontinuum laser source with linear polarization oriented parallel to the long wire axis. The two nanowire ends are visible as bright spots. This suggests an explanation by laser light exciting SPPs at the wire's ends, which are guided along the wire and are finally scattered into propagating light at the opposing ends. Between both ends no light is scattered from the wire. This is compatible with SPPs only coupling to propagating light at discontinuities, i.e. the ends of the wire. 
A direct proof of propagating SPPs in freely suspended silver nanowire is depicted in Figure~\ref{fig:opt}(c). The laser light from the supercontinuum source is now focused only onto one end of the wire. This can be seen as bright spot at the lower end of the wire. The interference fringes are due to an aperture effect by the trap electrodes.
The excited SPPs at the lower end of the wire now propagate over a distance of \unit{12}{\micro\meter} to the upper end. There, they are scattered into propagating light, which can be seen as a bright spot at the upper end of the wire in Figure~\ref{fig:opt}(c). In contrast, no scattered light can be observed when the laser is focused onto the wire at a position between both 
ends~\cite{Dickson2000, Sanders2006}. The polarization sensitive excitation and propagation is a clear signature for SPPs.

\begin{figure}
	\centering
		\includegraphics[width=0.8\linewidth]{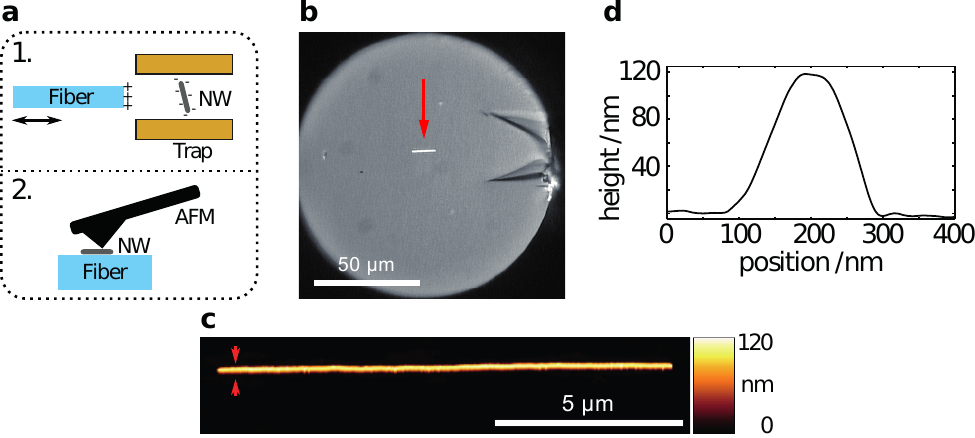}
		\caption{\textbf{Characterization of a deposited wire.}
		\textbf{a}, Charged nanowires from the trap can be deposited on a cleaved fiber with opposite surface charges. 
		Subsequently, the nanowire can be characterized by the atomic force microscope (AFM).
		\textbf{b}, Optical image of a silver nanowire extracted from the Paul trap.
		\textbf{c}, AFM image of the deposited wire. Using the AFM, the dimensions, especially the diameter of the nanowire can be measured with high accuracy.
		\textbf{d}, Cross section of the deposited nanowire measured with the AFM along mark indicated in c. The measured height corresponds to the diameter of the nanowire.
		}
		\label{fig:afm}
\end{figure}

After a first optical characterization of a wire, it can be removed from the trap for a precise analysis of its size and shape. In order to do this in a controlled way, we carefully insert a cleaved optical fiber along the axis at one end of the trap and approach it to the nanowire by a linear stage, as illustrated in Figure~\ref{fig:afm}(a). When the sign of the nanowire charge, set by the voltage polarity applied to the electrospray injector, is chosen opposite to the surface charges of the cleaved fiber, the nanowire will be attracted by the fiber and can be deposited on its facet~\cite{Kuhlicke2014}. After cleaving, the surface charge of the fiber is usually positive, but it changed by 
immersing the facet in an ethanol droplet with a high negative voltage from the electrospray injector applied.

After successful landing of a nanowire on the fiber facet the fiber is removed from the trap and installed in an atomic force microscope (AFM). Figure~\ref{fig:afm}(b) shows an optical image of a nanowire extracted from the Paul trap lying on a cleaved fiber facet. For further characterization, measurements with an atomic force microscope are carried out, as shown in Figure~\ref{fig:afm}(c). Such a measurement gives accurate information on the wire dimensions, especially its diameter, which is equal to the measured height of the wire (see Figure~\ref{fig:afm}(d)). A length of \unit{12}{\micro\meter} and a height of \unit{120}{\nano\meter} is measured for the shown nanowire. Controlled removal and AFM characterization is very reliable and can be performed on any of the trapped wires.

\begin{figure}
	\centering
		\includegraphics[width=\linewidth]{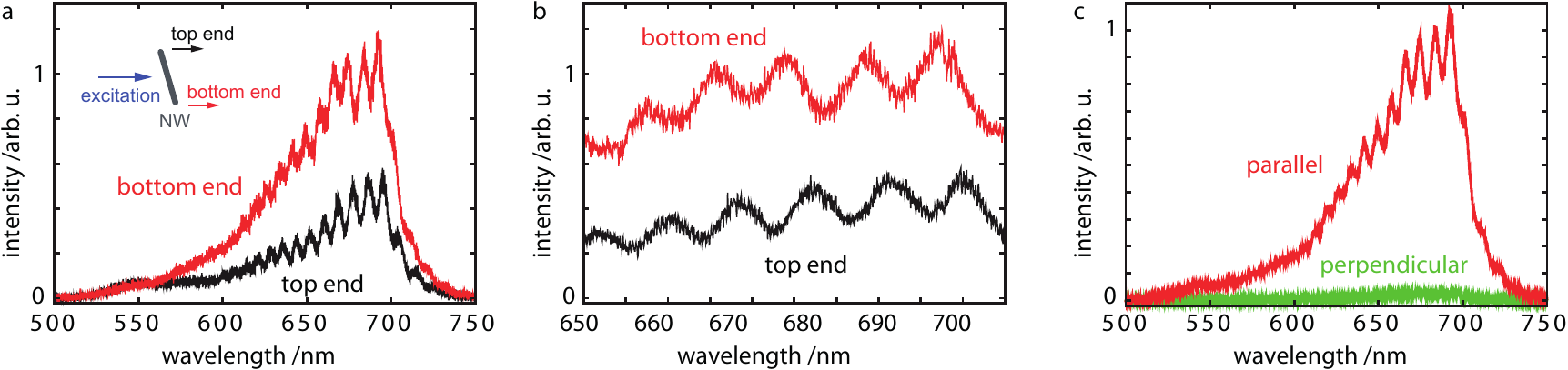}
		\caption{\textbf{White light scattering spectra acquired from levitating nanowires.}
		\textbf{a}, Optical spectra acquired from the top (black line) and the bottom ends (red line) of a silver nanowire approximately \unit{16}{\micro\meter} in length under illumination from the side using the supercontinuum source. The light is polarized parallel to the wire.
		\textbf{b}, Zoom into the region from \unit{650}{\nano\meter} to \unit{700}{\nano\meter}. relative shift of the modulation can be recognized between the spectra obtained from the two ends.
		\textbf{c}, Spectra of the same nanowire obtained from the bottom end with illumination polarized parallel (red line) and perpendicular 
		(green line) to the long wire axis. 
		}
		\label{fig:spectra}
\end{figure}

For a more detailed analysis of SPP propagation along suspended wires, spectra of locally scattered light are taken and investigated. In Figure~\ref{fig:spectra} the measured spectra of  light scattered from both ends of a \unit{16}{\micro\meter} long levitated wire can be seen. Excitation is done similar to the situation in Figure~\ref{fig:opt}(b) from the side with light from the supercontinuum source polarized parallel to the wire's axis. Scattered light is spatially selected via a pinhole. The spectra from either end of the wire clearly show a pronounced modulated structure. A much weaker signal and no modulation of the spectra is observed when the polarization of the excitation light is perpendicular to the long wire axis (Figure~\ref{fig:spectra}(c)).
A zoom into the region from \unit{650}{\nano\meter} to \unit{700}{\nano\meter} shown in Figure~\ref{fig:spectra}(b) reveals a relative shift of peaks in the spectra obtained from the two ends. Care has to be taken when interpreting this spectral structure, since there may be several contributions to the signal. At one end of the wire excitation light can be directly scattered into the detection path, but interference with light, which was coupled in at the opposite end and guided along the wire, may occur. If damping in the wire is low enough multiple reflections at the wire's ends may form a standing wave pattern similar to light confined in a Fabry-P\'{e}rot resonator~\cite{Ditlbacher2005,Kress2015}. 

\begin{figure}
	\centering
		\includegraphics[width=\linewidth]{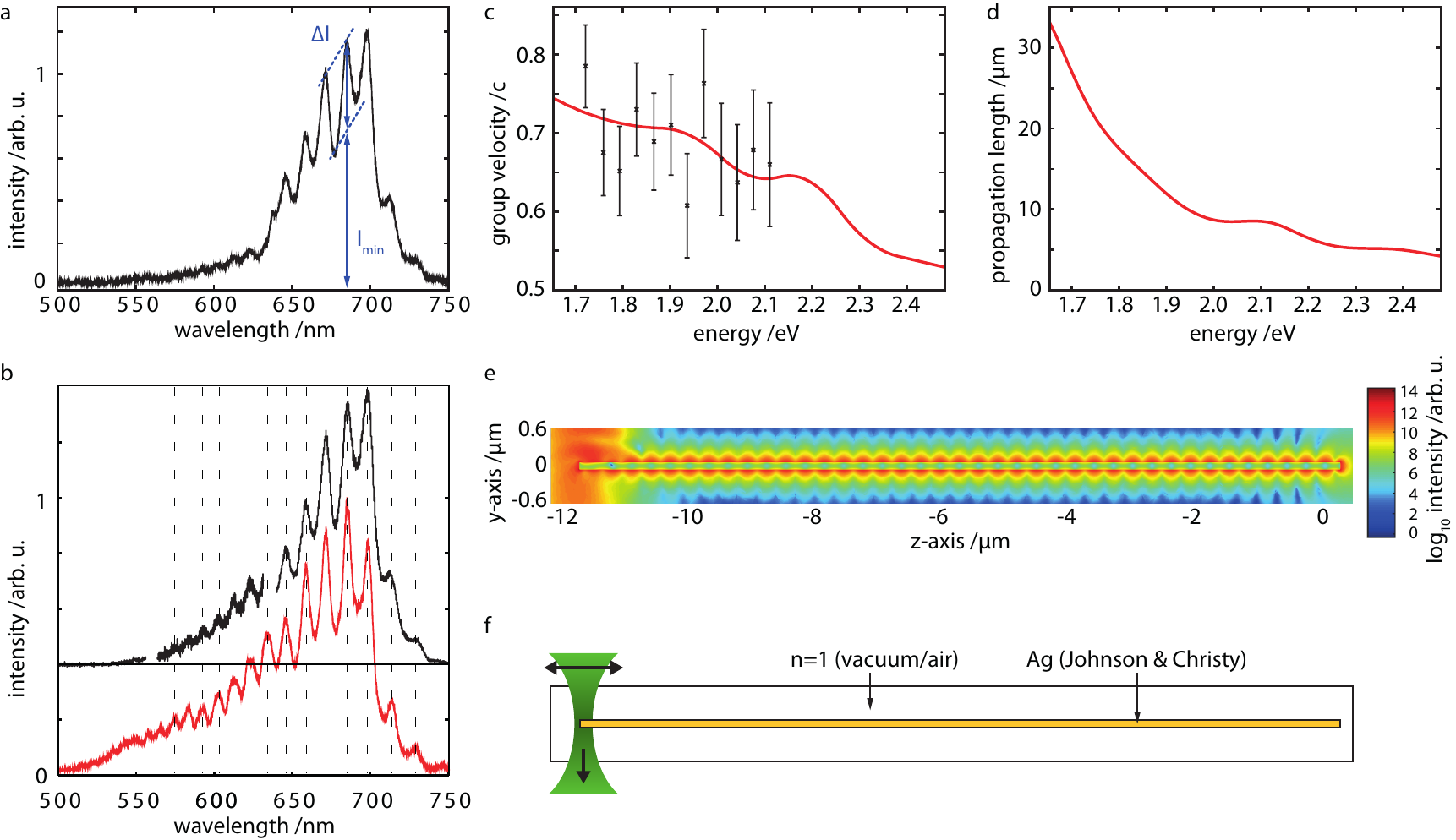}
		\caption{\textbf{White-light scattering spectrum of a single trapped nanowire in transmission and theoretical analysis.}
		\textbf{a}, Spectrum of a \unit{12}{\micro\meter} long nanowire measured using the supercontinuum source in transmission geometry. $I_{min}$ and $\Delta I$ indicate how the modulation depth is extracted from the spectra~\cite{Ditlbacher2005}. 
		The wire is the same as the one characterized in Figure~\ref{fig:afm}.
		\textbf{b}, Calculated spectrum according to a Fabry-P\'{e}rot model weighted with the spectrum of the supercontiunuum source (red line) compared to the experimental spectrum from (a) with a correction applied for the stepping artifact around \unit{600}{\nano\meter} (black line).
		\textbf{c}, Group velocity (normalized to the speed of light in vacuum $c$) of the SPPs calculated from the measured spectrum following Equation~\ref{eq:grvel} (black dots with the error bars given by the spectral resolution of the spectrometer (\unit{0.1}{\nano\meter})) compared to numerical simulations (red line).
	  	\textbf{d}, Propagation length of the SPPs as extracted from the numerical simulation.
		\textbf{e}, Two-dimensional projection of 3D simulation region and resulting electric field intensity (at a vacuum wavelength of  \unit{722.5}{\nano\meter}) when a Gaussian beam is focussed onto the end of the wire.
		\textbf{f}, Configuration of the simulation scheme.
		}
		\label{fig:sim}
\end{figure}

In order to clarify the observation another experiment was performed. Figure~\ref{fig:sim}(a) (black line) shows a spectrum acquired under illumination of only one end of the nanowire with white light from the supercontiumm source, similar to the situation in Figure~\ref{fig:opt}(c). Again, a pronounced peaked structure is observed. The nanowire shown is the same as the one characterized in Figure~\ref{fig:afm}, but due to the smaller length of the wire (\unit{12}{\micro\meter}), the distance of the peaks is larger compared with Figure~\ref{fig:spectra}. This is now a clean transmission experiment and the modulation can only be explained by formation of a standing wave pattern due to multiple reflections of guided SPP at the wire's ends. The number of SPP round trips, however, is limited by the propagation loss and reflectivity of the wire end facets. The large linewidth of the peaks indicates a plasmon Fabry-P\'{e}rot resonator with a rather low quality factor.

The simplicity of the experimental arrangement with a freely suspended wire together with a possibility to precisely measure the wire's size and shape (after deposition) allows for a direct comparison with theory with only two parameters, i.e. the complex effective refractive index $n_{\rm{eff}}$ of the wire and the reflectivity $R$ of its ends. We perform this analysis in the following.  

The roundtrip loss $r$ can be deduced from the relative modulation depth $\Delta I / I_{min}$ of the observed spectrum~\cite{Ditlbacher2005} by:
\begin{equation}
	\frac{\Delta I}{I_{min}} = \frac{4 r}{(1-r)^2} \,.
\end{equation}
Here $\Delta I$ and $I_{min}$ are the maximum and the minimum intensity at a certain wavelength approximated from neighboring maxima and minima, respectively, and  $r^2$ is the round-trip power attenuation factor \cite{Saleh1991}, where $r$ can be written as the product of losses due to the reflectivity $R$ and propagation loss $A$:
\begin{equation}
r = RA=R\rm{exp}(-\alpha \it{L}) \,.
		\label{eq:roundtrip}
\end{equation}
To be more precise $R$ is the average reflectivity of both wire ends ($R=\sqrt{R_1 R_2}$) while $\alpha$ is the attenuation constant with $\alpha$=4$\pi$ Im($n_{\rm{eff}}$)/$\lambda_0$ with the effective refractive index of the waveguide mode $n_{\rm{eff}}$ at the vacuum wavelength $\lambda_0$. Already at this point, a relative modulation depth of $0.53$ at a wavelength of \unit{685}{\nano\meter} can be derived directly from the measured spectrum in Figure~\ref{fig:sim}(a), which results in a value of $r=0.11$. Similar results are reported for experiments with silver nanowires on substrates~\cite{Allione2008,Shegai2011}. 

In order to decouple $R$ and $A$ as well as to simulate the experimental results we employ a simple Fabry-P\'{e}rot model (details see methods section). First, we use a finite element Maxwell's equations solver in the frequency domain to calculate the modes of an infinitely long wire. In the calculation the nanowire's diameter and its surrounding dielectric environment enter. The first one is precisely known from the AFM measurements, while the second one is trivially air. In order to resemble the real structure, the nanowire is assumed to consist of a silver nanowire with a diameter of $d=\unit{113}{\nano\meter}$, and a $t=\unit{3.5}{\nano\meter}$ thick shell of PVP that is used to stabilize the wires in solution (see methods section). 
The calculation yields the (wavelength-dependent) complex effective refractive index $n_{\rm{eff}}$. The effective index is then plugged into the formula for the normalized intensity in a Fabry-P\'{e}rot resonator~\cite{Saleh1991}:
\begin{equation}
	\frac{I(\lambda_0)}{I_{max}}=\frac{(1-R)^2}{|1-R \exp{(i\frac{4\pi L}{\lambda_0} n_{\rm{eff}})}|^2} \,. 
	\label{eq:fp2}
\end{equation}
with the cavity length $L$ (here the length of the wire) and the reflectivity $R$ as defined in equation $\ref{eq:roundtrip}$. The intensity $I$ is proportional to the transmission $T$ through the cavity and thus to the measured scattering spectrum. Keep in mind, that $n_{\rm{eff}}$ is a complex number. Finally, the calculated spectrum is weighted with the spectrum of the white light source used for excitation. The reflectivity $R$ remains either as a fit parameter or can be calculated accordingly.

Figure~\ref{fig:sim}(b) (red line) shows the results of the calculation in comparison to the measurements (black line) with the \unit{12}{\micro\meter} long nanowire. A (wavelength independent) reflectivity of $R = 0.3$ is assumed. The agreement is excellent. Note, that not only the peak positions in the measured spectrum, but also the widths are nicely reproduced, indicating that the spectral response of the levitating nanowire can be accurately described and fully understood as a simple single-mode Fabry-P\'{e}rot resonator and modeled (quantitatively). We conclude that basically any waveguide-based nanoresonator can thus be described with this most simple approach, which should be especially interesting for hybrid systems as studied for strong coupling or nanolasing. 

From the calculated effective refractive index also the group velocity (see Figure~\ref{fig:sim}(c)) as well as the propagation length of the SPP (see Figure~\ref{fig:sim}(d)) are obtained. The former is in agreement with values estimated directly from the measured spectra with an independent method (see Figure~\ref{fig:sim}(c) and methods section). 

Finally, we performed a full 3D simulation (see methods section) to verify the results of the simple Fabry-P\'{e}rot model. The geometry of this 3D simulation (Figure~\ref{fig:sim}(f)) fully matches the experiment with the source being a Gaussian beam. The wavelength is chosen to be at a vacuum wavelength of \unit{722.5}{\nano\meter} corresponding to a Fabry-P\'{e}rot resonance. One can clearly identify a standing wave pattern along the nanowire with 40 maxima, which results in a surface plasmon wavelength of about \unit{600}{\nano\meter} and thus an effective refractive index of $n_{\rm{eff}}=1.20$. This is in perfect agreement with the computed effective refractive index of $n_{\rm{eff}}=1.19$, calculated for the infinite wire with the finite element Maxwell's equation solver at the same frequency.

In conclusion, we measured and calculated the plasmonic properties of a silver nanowire levitated in a Paul trap. Focussed excitation of SPP on one end and collection of scattered light on the other end resembles an ideal scenario of a transmission measurement. Without any distortion from supporting structures the observed SPP propagation perfectly agrees with numerical calculation, where a simple and intuitive model is valid. This insight should motivate researchers to describe also finite nanowaveguides on substrates based on such a simple model which might be especially intuitive for complex hybrid systems for nanolasing or strong coupling. 
Not only SPP propagation, but also other features of plasmonic particles can be studied in our experiment. The perfect thermal isolation suggests investigation of heat generation or transport on the nanoscale~\cite{Brongersma2015}. Furthermore, hot electrons and plasmonic effects for the enhancement of catalytic reactivity could be studied in a Paul trap~\cite{Kim2016}. To this end plasmonic particles could be trapped in gases that are reactants of the catalytic reaction under consideration. Chemical products could than be detected with gas sensors or sensitive spectroscopic methods.
Since nearly any object dispersed in solution can be trapped in a Paul trap, our experiment opens the way to levitate more 'fragile' objects, such as graphene sheets, flakes of two-dimensional materials or even droplets of liquids or molten metals, the latter being a hardly explored field in plasmonics~\cite{Kuhlicke2013}.  
Finally, trapped particles can be coupled with other simultaneously trapped objects, such as solid-state quantum emitters~\cite{Kuhlicke2014} in order to explore quantum features of coupled systems~\cite{Kuhlicke2015}.

\section*{Acknowledgement}
Funding by DFG (Sfb951, project B2) and the Einstein Foundation Berlin 
is acknowledged. O.B. thanks CNPq Brasil (science without borders program). A.W.S. thanks JSPS
for the fellowship for overseas researchers. The authors thank JCMwave for support.

\section*{Methods}
\textbf{Paul trap} \\
A linear Paul trap with end-cap electrodes is used to levitate the charged particles in the experiment, as described before~\cite{Kuhlicke2014}. Four brass electrodes with a total length of \unit{7}{\milli\meter} and an outer diameter of \unit{4}{\milli\meter} form the trap with an inner free space radius of \unit{1.75}{\milli\meter}. The electrodes are held in place by milled fiberglass boards. In the radial trap direction particles are confined in an oscillating electric quadrupole potential. Therefore, one pair of diagonally opposing electrodes is connected to an AC high voltage, while the other pair is held at ground. Axial stabilization is achieved by additional end-cap electrodes on both sides of the trap. The end-cap electrodes, which are connected to individual static potentials, are located at the ends of each of the four quadrupole electrodes. This configuration creates an adjustable static potential in the axial trap direction, enabling axial confinement and adjustment of the particle position in the trap, without blocking the axial access to the trap center, which is used for illumination and insertion of cleaved optical fibers for particle deposition. A commercial frequency generator in combination with a homemade amplifier~\cite{kuhlicke_broadband_2014} are used to generate the AC high voltage with an amplitude of up to \unit{1.7}{\kilo\volt} within a frequency range from \unit{20}{\hertz} to \unit{100}{\kilo\hertz}. Static voltages up to \unit{120}{\volt} are created by a commercial laboratory power supply and applied to the end-cap electrodes through a switch-box with adjustable outputs for each connection line.

\textbf{Optical setup} \\
The trap is integrated in a homemade optical microscope for particle observation. A  microscope objective (Olympus, LMPLFLN 50x) with a numerical aperture of $0.5$ and a long working distance of \unit{10.6}{\milli\meter} is adjusted to the center of the trap, perpendicular to the trap axis. Fine-adjustment of the trap and the objective positions is done by linear translation stages and piezoelectric actuators. A supercontiuum fiber laser (PicoQuant, Solea, SCS in Figure~\ref{fig:setup}) is used to illuminate and excite the particles in the trap, either strongly focused through the microscope objective or only slightly focused along the trap axis. The fiber laser can be used as a white-light source with a spectral range from \unit{525}{\nano\meter} to \unit{700}{\nano\meter}, or at an arbitrary wavelength within this spectral range with a width of \unit{5}{\nano\meter} limited by the integrated band-pass filter. Additionally, white light from a halogen lamp can also be used for illumination from above the trap. For detection, the light scattered from the particle and collected by the microscope objective is spatially selected by a pinhole (Spatial filter in Figure~\ref{fig:setup}) and sent either to a CCD camera or to a spectrometer. The different excitation and detection paths are switched by flip mirrors. In front of the objective, a beam sampler is used to separate the incoming and the outgoing light paths in such a way that only 10 \% of the excitation light is reflected into the objective and nearly 90 \% of the collected light is sent towards detection. The axis of the linear polarized laser emission can be rotated by a half-wave plate ($\lambda / 2$ in Figure~\ref{fig:setup}) in the excitation path, but is set by a subsequent linear polarizer (Pol. in Figure~\ref{fig:setup}). This combination is used to adjust the laser intensity for a chosen linear polarization orientation. A second linear polarizer (Pol. in Figure~\ref{fig:setup}) with orientation orthogonal to the first one can be used to block the excitation light in the detection path.

\textbf{Numerical Simulations} \\
The plasmonic Fabry-P\'{e}rot resonator is modeled with a finite element Maxwell's equations solver (JCMwave) in the frequency domain. In all calculations the dielectric permittivity values measured by Johnson and Christy~\cite{Johnson1972} are used. The metallic nanowire is assumed to be capped by a thin shell of a stabilizing polymer (the sample used in the experiments is stabilized by PVP (PlasmaChem)) of \unit{3.5}{\nano\meter} with refractive index of 1.4. A nanowire diameter $d$ of \unit{120}{\nano\meter} and a length $L$ of \unit{12}{\micro\meter} as measured with the AFM is used in the numerical simulations.

With a propagating mode solver, the guided modes of an infinitely long cylinder are computed. The eigenvalues of this computation are the complex effective refractive indices $n_{\rm{eff}} = k_{spp}/k_0$ with the SPP's wave vector $k_{spp}$ and the photon wave vector in vacuum or air $k_0$. The real and imaginary part of $n_{\rm{eff}}$ were calculated as a function of vacuum wavelength $\lambda_0$ together with the corresponding group velocity ($v_{gr} = \frac{d\omega}{dk}$) and the mean propagation length ($L_{prop} = \frac{\lambda_0}{4\pi Im(n_{\rm{eff}})}$). In order to compare the calculated values for the group velocity with values derived from the experiment we used the estimation~\cite{Allione2008}:
\begin{equation}
	v_{gr}=2 L c \frac{\Delta\lambda}{\lambda_0^2}
	\label{eq:grvel}
\end{equation}
which yields the group velocity at the center wavelength $\lambda_0$ between two adjacent peaks of the measured scattering spectrum. Figure~\ref{fig:sim}(c) shows these group velocity calculated from the spectrum shown in Figure~\ref{fig:sim}(a).
Group velocities of about 0.7 times the vacuum speed of light are found, which is in good agreement with other reported experimental and theoretical values.

In addition to  the fundamental mode, two degenerated loosely guided modes with $n_{\rm{eff}} \approx 1$ are found. Excitations of these modes were ignored in the analysis as they experience negligible reflection at the end-facets of the nanowire, as was tested in full three-dimensional numerical simulations with the guided modes acting as light source. 

The results of the simple Fabry-P\'{e}rot model have been checked with an independent simulation approach like the one shown in Figure~\ref{fig:sim}(e) and (f). In these simulations a full 3D description of the problem was made, using either a dipolar light source or a Gaussian beam (composed by a set of plane waves, yielding a Rayleigh length of \unit{2}{\micro\meter}). The computed field pattern showed standing waves that are in agreement with the predicted plasmon wavelength computed with the propagating mode solver. Further, emission spectra from the distal end were recorded by monitoring the electromagnetic flux through the right side of the computational domain. These spectra showed the same periodicity as the result from the Fabry-P\'{e}rot model. Note, that modeling with the Fabry-P\'{e}rot model is significantly faster since it is based on quickly computable propagating mode solutions.

\section*{Author contributions}
A.S. and A.K. performed experiments and measurements; G.K. performed numerical simulations; all authors contributed to data interpretation, writing and revision of the manuscript.

\section*{Competing financial interests}
The authors declare no competing financial interests.

\section*{Supporting Information}
Supporting information on the influence of the cross-section of the nanowires is available.

\providecommand{\latin}[1]{#1}
\makeatletter
\providecommand{\doi}
  {\begingroup\let\do\@makeother\dospecials
  \catcode`\{=1 \catcode`\}=2\doi@aux}
\providecommand{\doi@aux}[1]{\endgroup\texttt{#1}}
\makeatother
\providecommand*\mcitethebibliography{\thebibliography}
\csname @ifundefined\endcsname{endmcitethebibliography}
  {\let\endmcitethebibliography\endthebibliography}{}

\end{document}